\documentclass{webofc}
\usepackage[varg]{txfonts}   
\begin{document}
\title{Hadron Spectroscopy with lattice QCD: challenges and opportunities}
%
%

\author{\firstname{John} \lastname{Bulava}\inst{1}\fnsep\thanks{\email{john.bulava@ruhr-uni-bochum.de}}
}

\institute{Fakult\"{a}t f\"{u}r Physik und Astronomie, Institut f\"{u}r Theoretische Physik II, Ruhr-Universit\"{a}t Bochum, 44780 Bochum, Germany}

\abstract{%
	Ongoing challenges in computing the spectrum of hadronic resonances and shallow bound-states from lattice QCD are reviewed. Since such states  
 are identified as poles in the scattering matrix, nearby non-analyticities must be treated to analytically continue to complex center-of-mass energies. Significant lattice spacing effects have also been observed in some channels, necessitating a continuum limit.  
Recent achievements are also highlighted, including lattice investigations of states in the charm region, baryon-baryon scattering, and the 
first coupled channel meson-baryon amplitude in the $\Lambda(1405)$ channel. 
}

\maketitle

The spectrum of hadrons is a rich manifestation of the complex strong dynamics of QCD. In addition to conventional mesons and baryons, more exotic color-singlets are possible, such as tetra- and penta-quarks, or configurations with constituent gluons like hybrids and glueballs.  
The nature of the atomic nucleus suggests additional possibilities. Like the van der Waals forces between atoms, residual interactions between hadrons could generate `molecular' bound states akin to conventional nuclei but with meson or hyperon constituents.  

Much experimental and theoretical effort is oriented in this direction, 
resulting in increasing evidence for many exotic states~\cite{Chen:2022asf,Richard:2016eis,Briceno:2015rlt}. 
The exploration of these scenarios requires the study of resonances and near-threshold bound states, which are identified as poles in the scattering matrix analytically continued to complex center-of-mass energies. Lattice QCD can contribute (in principle) not only by computing scattering amplitudes at the physical point to match experiment, but also by tracing the behaviour of states toward more tractable limits of QCD. These limits include the isospin limit where $m_{\rm u} = m_{\rm d}$ and electroweak interactions are absent, the chiral limit $m_{\rm u,d} \rightarrow 0$, the heavy-quark limit $m_{\rm b} \rightarrow \infty$, 
the $SU(3)$-flavor limit with $m_{\rm u} = m_{\rm d} = m_{\rm s}$, and the large-$N_{\rm color}$ or large-$N_{\rm flavor}$ limits.  

However, the first-principles computation 
of scattering amplitudes from lattice QCD is considerably more difficult than for stable single-hadron states. Higher $n$-point Euclidean correlation functions are required, which suffer from a computational cost which increases with $n$ and an exponentially decreasing signal-to-noise ratio~\cite{Beane:2010em}. Furthermore, a matrix of correlation functions is required between interpolating operators for all relevant finite-volume states. In addition to improvements in available computing power, these difficulties have been ameliorated by efficient algorithms for quark propagation~\cite{Peardon:2009gh,Morningstar:2011ka}, such that precise determination of the finite-volume spectrum in many channels is possible. These algorithms enable a factorization of the quark propagator and reduce the construction of correlation functions to tensor contraction operations, facilitating the computation of correlation matrices between many interpolating operators. 

\begin{figure}
	\includegraphics[width=\textwidth]{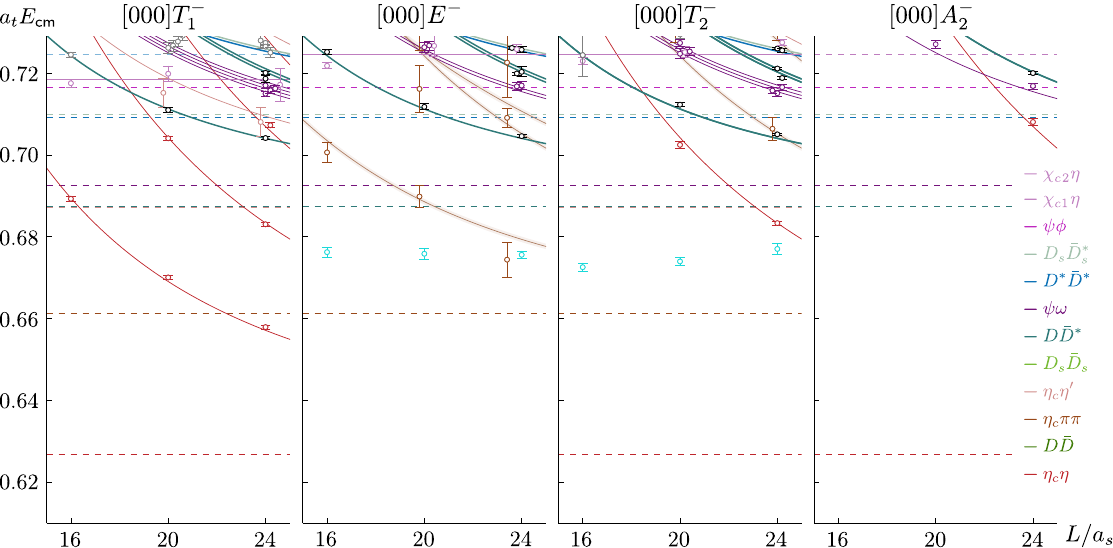}
	\caption{\label{f:hsc} Taken from Refs.~\cite{Wilson:2023hzu,Wilson:2023anv} by the HadSpec collaboration, the finite-volume spectrum of charmonium states (neglecting charm annihilation) in some irreps with zero total momentum computed on a 
	single ensemble of $N_{\rm f}=2+1$ gauge configurations with $m_{\pi} = 391\,{\rm MeV}$. Three different spatial volumes are employed with $L/a_{\rm s} = 16,\, 20,\,24$ as denoted on the horizontal axis. The horizontal dashed lines indicate various thresholds, as denoted in the legend, while solid curved lines correspond to non-interacting finite-volume energies. The band of states in the $E^{-}$ and $T_2^{-}$ irreps indicates the presence of a tensor resonance. }
\end{figure}

A recent example demonstrating the efficacy of such methods is given by the HadSpec collaboration in 
Refs.~\cite{Wilson:2023hzu,Wilson:2023anv}, which computes the finite volume charmonium spectrum (neglecting charm-quark annihilation) on a single ensemble of $N_{\rm f}=2+1$ gauge configurations with $m_{\pi} = 391\,{\rm MeV}$. This comprehensive computation includes interpolators for all two-hadron channels containing a charm and anti-charm quark below $4100{\rm MeV}$, in addition to some three-meson operators.  The finite volume spectra from Refs.~\cite{Wilson:2023hzu,Wilson:2023anv} for some irreps at total zero momentum are shown in Fig.~\ref{f:hsc}. A list of the kinematically relevant channels, all of which require corresponding interpolators, is also given there.  

We turn now to the determination of scattering amplitudes from finite-volume spectra like the one in Fig.~\ref{f:hsc}. 
Due to the Euclidean metric signature in lattice QCD, the Haag-Ruelle asymptotic formalism familiar to Minkowski space~\cite{Haag,Ruelle} cannot be naively applied: the large time-separation limit of Euclidean correlators does not generally yield on-shell scattering amplitudes~\cite{Maiani:1990ca}.  
Fortunately, a work-around was developed by L\"{u}scher~\cite{Luscher:1990ux} 
which relates 
the finite-volume spectrum of two-hadron states below $n\ge 3$ hadron 
thresholds to the two-to-two scattering amplitude. As demonstrated above, the finite-volume spectrum 
is directly computable from the large-time limit of Euclidean correlators in lattice QCD using a variational approach~\cite{Michael:1982gb}. In addition to two-to-two scattering, considerable progress has been made on three-hadron amplitudes~\cite{Romero-Lopez:2021zdo, Dawid:2023lcq}. A recent review of the finite-volume formalism is given in Ref.~\cite{Hansen:2022csj}.

The relation between two-to-two scattering amplitudes and the finite-volume spectrum below $n\ge 3$ hadron thresholds is given by~\cite{Morningstar:2017spu}
\begin{equation}\label{e:fv}
	{\rm det}[ K^{-1}(E_{\rm cm}^{\rm FV}) - B_L(\boldsymbol{P})]
\end{equation}
where $K = (2T^{-1}+i)^{-1}$ is related to the scattering $T$-matrix in the usual way, and $B_L$ encodes the effect of the finite volume. The determinant is taken over all open scattering channels and partial waves, and is typically evaluated in a basis which is block-diagonal in the irreducible representations of the little group for total momentum $\boldsymbol{P}$. This infinite-dimensional relation must be truncated at some maximum angular momentum $\ell_{\rm max}$, and holds up to corrections which are suppressed exponentially with the spatial extent $L$. 

It should be emphasized that Eq.~\ref{e:fv} holds also for finite-volume energies \emph{below} the two-particle scattering threshold, which occur for an sufficiently attractive interaction. This uniquely provides direct constraints on the scattering amplitude at complex center-of-mass momenta and aides the determination of near-threshold bound-state poles. However, as first pointed out in Ref.~\cite{Green:2021qol} for $\Lambda\Lambda$ scattering, levels significantly below threshold may overlap left-hand (cross-channel) cuts, which are not accommodated by Eq.~\ref{e:fv}. This is not purely an academic observation: Ref.~\cite{Green:2021qol} determined a number of finite-volume levels below the left hand cut but could not use them to constrain the amplitude since Eq.~\ref{e:fv} is inapplicable.

\begin{figure}
\centering 
	\includegraphics[width=\textwidth]{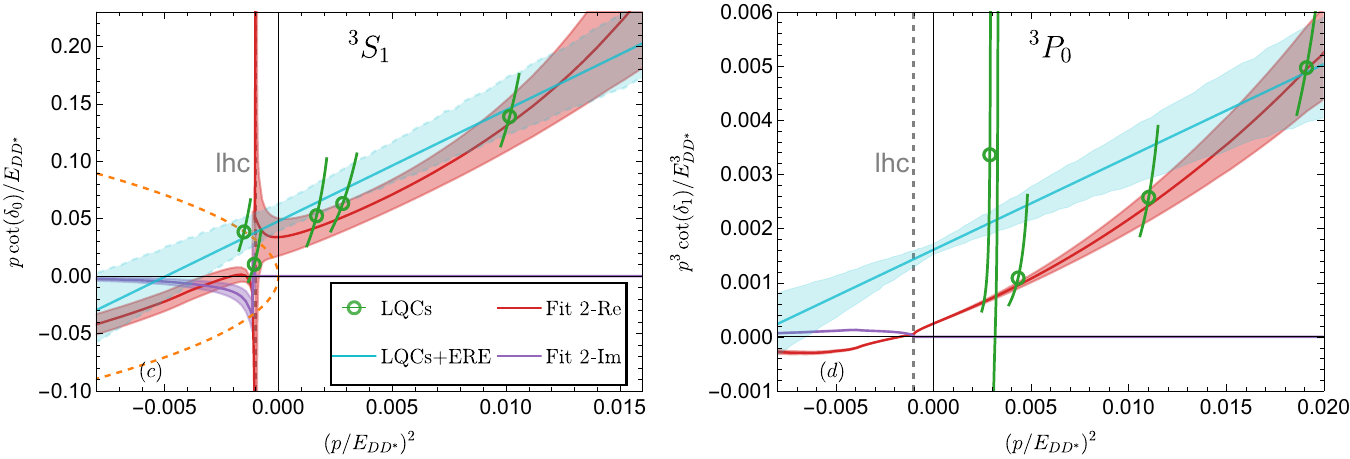}
	\caption{\label{fig:rub} The $^3 S_{1}$ and $^{3} P_0$ partial waves in $DD^{*}$ scattering from Ref.~\cite{Meng:2023bmz}. Shown are fits to the finite-volume $DD^{*}$ energy levels of Ref.~\cite{Padmanath:2022cvl} computed on a single $N_{\rm f} = 2+1$ ensemble with $m_{\pi}=280\,{\rm MeV}$. The points are obtained from Eq.~\ref{e:fv} in the leading partial wave approximation, and labelled `LQCs', while a fit to them using Eq.~\ref{e:ere} is shown in blue and labelled `LQCs+ERE'. The real and imaginary parts of fits including one-pion exchange are shown in red (labelled `Fit 2-Re') and blue (`Fit 2-Im'), respectively. The dashed lines are $ip$ and intersect the fit at a (virtual) bound state pole. The position of the left-hand cut is given by the vertical lines labelled `lhc'. }
\end{figure}

Theoretical effort is underway to augment the finite-volume formalism to handle the closest left-hand cut which, for two heavy scattering particles like nucleons, is due to the $t$- and $u$-channel exchanges of a single pion. Ref.~\cite{Raposo:2023oru} replaces Eq.~\ref{e:fv} by one constraining an alternative infinite-volume quantity $\overline{\mathcal{K}}^{\rm os}$ and a coupling $g$. These intermediate objects are then related to the desired amplitude by an integral equation. By contrast, Ref.~\cite{Meng:2023bmz} uses an effective field theory approach with explicit one-pion exchange and solves an eigenvalue problem to obtain levels both above and below the left hand cut. It is based on earlier work which treats Eq.~\ref{e:fv} in the plane-wave basis~\cite{Meng:2021uhz}, and a previous EFT treatment of the same system~\cite{Du:2023hlu}. As a first application, Ref.~\cite{Meng:2023bmz} incorporates 
the leading left-hand cut due to one-pion exchange in an analysis of the lattice QCD energy levels for $D^{*}D$ scattering from Ref.~\cite{Padmanath:2022cvl}, which are at quark masses such that $m_{\pi}=280\,{\rm MeV}$ and the $D^{*}$ is stable. This application is summarized in Fig.~\ref{fig:rub}, which compares 
fits to  the
next-to-leading order (NLO) effective range parametrization 
\begin{equation}\label{e:ere}
	p_{\rm cm}^{2\ell+1} \, {\cot} \,\delta_{\ell} = \frac{1}{a_{\ell}} + \frac{r}{2} p_{\rm cm}^2   
\end{equation}
to points obtained from Eq.~\ref{e:fv} treating only the lowest contributing partial wave against another fit including one-pion exchange. Both fits indicate the presence of a (virtual) bound state pole.  

\begin{figure}
	\centering
  \includegraphics[width=0.75\textwidth]{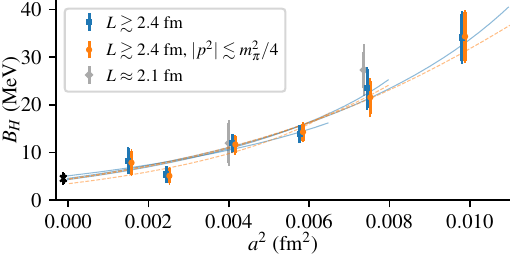}
	\caption{\label{f:mainz} From Ref.~\cite{Green:2021qol}, the binding energy of the $H$ dibaryon at $m_{\pi} = m_{\rm K} = m_{\eta} = 420\, {\rm MeV}$ determined from ensembles at six different lattice spacings. The different points at each lattice spacing correspond to various data selection procedures performed in the analysis. }
\end{figure}

In addition to left-hand cuts, another source of systematic error which has been studied recently is the finite lattice spacing. In practice it is difficult to engineer gauge field ensembles at different lattice spacings with the same physical volume, hampering a direct continuum limit of the finite-volume spectrum. An alternative is therefore to extract a pole position at each lattice spacing separately and perform a continuum extrapolation of the results. This strategy was employed in Ref.~\cite{Green:2021qol} to examine the continuum limit of the $H$ dibaryon binding energy at the $SU(3)$ flavor-symmetric point $m_{\pi} = m_{\rm K} = m_{\eta} = 420{\rm MeV}$. The results are shown in Fig.~\ref{f:mainz} and reveal significant cutoff effects in the 
extracted pole position. 

To recap, the workflow for determining the pole positions of resonances and bound states from lattice QCD simulations is as follows. Given an ensemble of gauge configurations, determine the correlation functions between all relevant two-hadron and single-hadron interpolating operators. From these correlation functions, the finite-volume spectrum is determined as in Fig.~\ref{f:hsc}. Then, Eq.~\ref{e:fv} is employed to constrain parametrization of the scattering amplitude using the finite-volume spectrum. Finally, these parametrizations are used to analytically continue to the complex plane and search for poles. This last step is illustrated in Refs.~\cite{BaryonScatteringBaSc:2023ori,BaryonScatteringBaSc:2023zvt} which compute the coupled-channel $I=0$, strangeness $S=1$, $\pi\Sigma-\bar{K}N$ scattering amplitude relevant for the $\Lambda(1405)$ resonance. This analysis performed on single ensemble of $N_{\rm f}=2+1$ gauge configurations with $m_{\pi} = 200\,{\rm MeV}$ and lattice spacing $a = 0.065\,{\rm fm}$. The determination of the finite volume spectrum requires the algorithms for all-to-all propoagation discussed above to determine correlation matrices between $\pi\Sigma$, $\bar{K}{N}$, and single-baryon interpolators at various total momenta. The finite volume spectrum is used to constrain various parametrizations of the amplitude similar to Eq.~\ref{e:ere}. The nearest left-hand cut for this system is far from the region of interest, and is therefore ignored.
\begin{figure}
	\centering
	\includegraphics[width=0.75\textwidth]{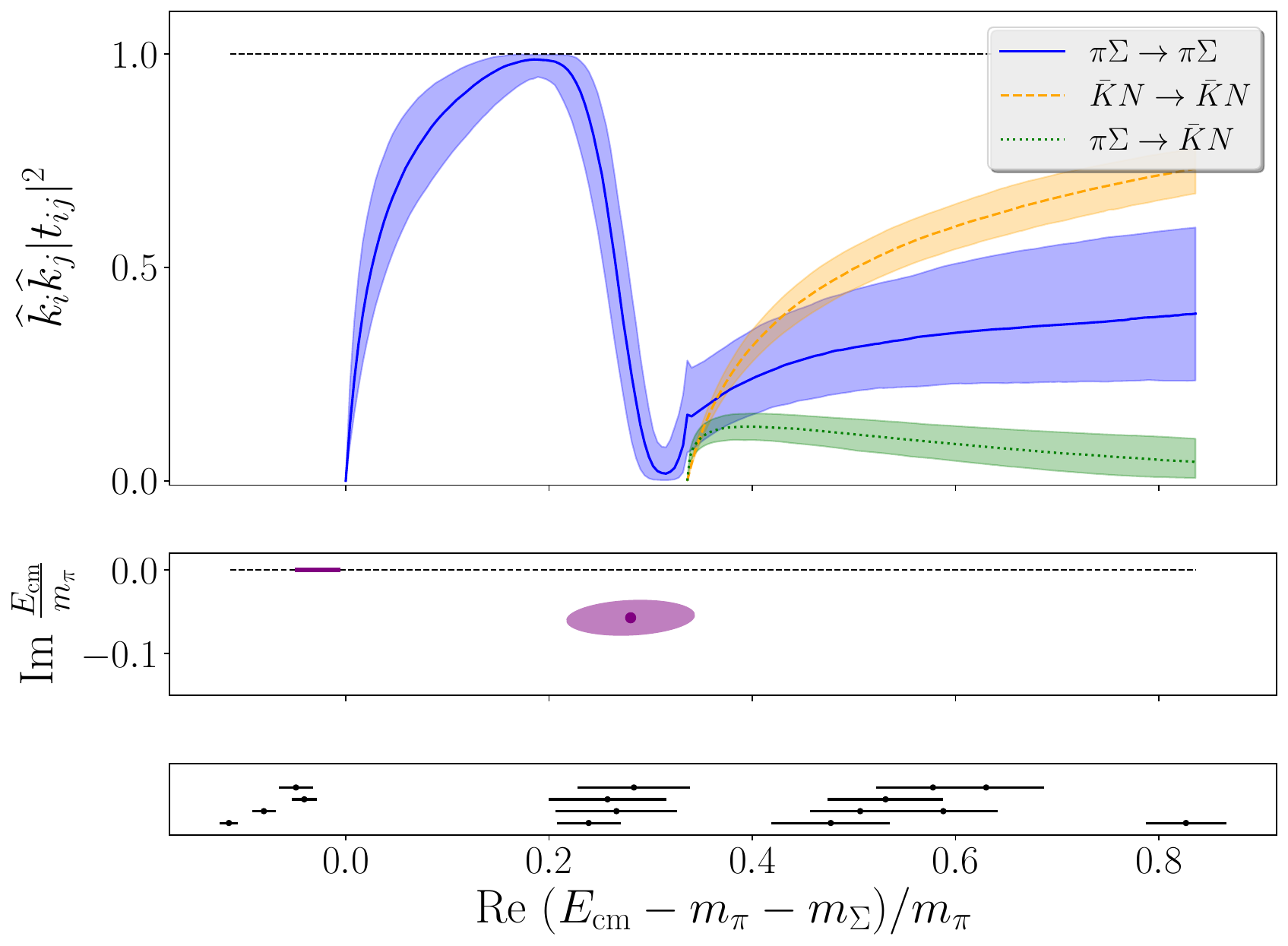} 
	\caption{\label{f:lambda} From Refs.~\cite{BaryonScatteringBaSc:2023ori,BaryonScatteringBaSc:2023zvt}, the coupled-channel $\pi\Sigma-\bar{K}N$ scattering amplitude computed on a single $N_{\rm f} = 2+1$ flavor ensemble with $m_{\pi}=200\,{\rm MeV}$ and lattice spacing $a = 0.065\,{\rm fm}$. The finite volume spectrum is shown in the bottom panel, and the amplitude with statistical errors in the top panel. The center panel shows the pole positions found in the analytic continuation of the amplitude, indicating a virtual bound state below $\pi\Sigma$ threshold and a resonance just below $\bar{K}{N}$ threshold.} 
\end{figure}
Parametrization constrained by the spectrum are then analytically continued to the complex plane, resulting in the two poles shown in Fig.~\ref{f:lambda}. This first lattice QCD analysis of the coupled-channel $\pi\Sigma$-$\bar{K}N$ amplitude clearly yields two poles in the $\Lambda(1405)$ system, albeit at the unphysically large pion mass of $m_{\pi}=200{\rm MeV}$. The pole positions determined here are consistent with the recent $SU(3)$ chiral effective theory analysis of Ref.~\cite{Guo:2023wes}. Such approaches typically predict two poles, generated by the flavor-$SU(3)$ octet and singlet channels. Future lattice computations at the $SU(3)$ point employed by Ref.~\cite{Green:2021qol} could test this hypothesis. Furthermore, the methods employed here are directly applicable at the physical point to compare with the recent (preliminary) experimental analysis of the Glue-X collaboration~\cite{Wickramaarachchi:2022mhi}. Cutoff effects in the pole positions must also be investigated for quantitative accuracy.  

In summary, the study of near-threshold bound states in lattice QCD has reached a point where several new sources of systematic error must be considered, including the influence of left-hand cuts and lattice spacing effects. The finite-volume formalism is well-equipped to study these systems since finite-volume energy levels below threshold provide direct constraints on the amplitude in the 
complex plane. The ongoing theoretical developments described here to treat the leading left-hand cuts due to single-particle exchange will be incorporated in the analysis of numerical data, and systems relevant for experiment are increasingly studied near or at the physical point. Another interesting development not reviewed here concerns the determination of spectral densities 
from Euclidean correlators. As reviewed in Ref.~\cite{Bulava:2023mjc}, this approach has enabled the determination of inclusive rates from lattice QCD for the first time. A formalism for extending this to exclusive amplitudes has also been proposed~\cite{Bulava:2019kbi}, but has not yet been 
applied. 

\begin{acknowledgement}
	We thank the organizers for an enjoyable MENU 2023 conference, as well as the authors of Refs.~\cite{Raposo:2023oru} and~\cite{Meng:2023bmz} for discussions about their work. 
\end{acknowledgement}

\bibliography{latticen}

%
%

\end{document}